\documentclass[usenatbib,onecolumn]{mn2e}

\def\gcc{\hbox{\rm\hskip.35em  g cm}$^{-3}$}

\def\eg{{\it e.g.}}
\def\lap{\hbox{${_{\displaystyle<}\atop^{\displaystyle\sim}}$}}
\def\gap{\hbox{${_{\displaystyle>}\atop^{\displaystyle\sim}}$}}

\def\eg{{\it e.g.}}
\def\lap{\hbox{${_{\displaystyle<}\atop^{\displaystyle\sim}}$}}
\def\gap{\hbox{${_{\displaystyle>}\atop^{\displaystyle\sim}}$}}

\usepackage{graphicx}
\usepackage{subfigure}

\title[Origin of Magnetar Flares]
{Constraining the Origin of Magnetar Flares}
\author[B. Link]{Bennett Link\thanks{E-mail:
link@physics.montana.edu}\\
Department of Physics, Montana State University, Bozeman, Montana, 
59717, USA \\
Department of Physics, Monash University, Melbourne, Victoria 3800, Australia
}
\begin{document}

\date{\today}

\pagerange{\pageref{firstpage}--\pageref{lastpage}} \pubyear{2013}

\maketitle

\label{firstpage}

\begin{abstract}

Sudden relaxation of the magnetic field in the core of a magnetar
produces mechanical energy primarily in the form of shear waves which
propagate to the surface and enter the magnetosphere as relativistic
Alfv\'en waves. Due to a strong impedance mismatch, shear waves
excited in the star suffer many reflections before exiting the
star. If mechanical energy is deposited in the core and is converted
{\em directly} to radiation upon propagation to the surface, the rise
time of the emission is at least seconds to minutes, and probably
minutes to hours for a realistic magnetic field geometry, at odds with
observed rise times of $\lap 10$ ms for both and 
giant flares. Mechanisms for both small and giant flares that rely on
the sudden relaxation of the magnetic field of the core are rendered
unviable by the impedance mismatch, requiring the energy that drives
these events to be stored in the magnetosphere just before the flare.
A corollary to this conclusion is that if the quasi-periodic
oscillations (QPOs) seen in giant flares represent stellar
oscillations, they must be excited {\em by the magnetosphere}, not by
mechanical energy released inside the star. Excitation of stellar
oscillations by relativistic Alfv\'en waves in the magnetosphere could
be quick enough to excite stellar modes well before a giant flare
ends, unless the waves are quickly damped.

\end{abstract}


\section{Introduction}
\maketitle

Soft-gamma repeaters (SGRs) are strongly-magnetized neutron stars with
magnetic fields of $B=10^{14}-10^{15}$ G that produce frequent,
short-duration bursts ($\lap 1$ s) of $\lap 10^{41}$ ergs in hard
x-ray and soft gamma-rays, with the peak luminosity in the burst
typically being reached in under 10 ms (\eg,
\citealt{wt06,mereghetti08}).  SGRs occasionally
produce giant flares that last $\sim 100$ s; the first giant flare to
be detected occurred in SGR 0526-66 on 5 March, 1979
\citep{barat_etal79,mazets_etal79,cline_etal80}, releasing 
$\gap 6\times 10^{44}$ erg \citep{evans_etal80}, and rising to near its
peak luminosity in $<2$ ms
\citep{mazets_etal79,hurley_etal99}. The duration of the initial
bright peak was 0.1-0.2 s \cite{mazets81}.  The August 27th 1998 giant
flare from SGR 1900+14 liberated $\gap 4\times 10^{43}$ erg, with a
rise time of $<4$ ms \citep{hurley_etal99,feroci_etal99}. The duration
of the initial peak was $\sim 1$ s \citep{hurley_etal99}. On December
27, 2004, SGR 1806-20 produced the largest flare yet recorded, with a
total energy yield of $\gap 4\times 10^{46}$ ergs and a rise time of
$<1$ ms
\citep{hurley_etal05,palmer_etal05,terasawa_etal05}. The duration of
the initial peak was $\sim 0.2$ s \citep{hurley_etal05}. All of these
energy estimates assume isotropic emission.

The giant flares in SGR 1900+14 (hereafter SGR 1900) and SGR 1806-20
(hereafter SGR 1806) showed rotationally phase-dependent,
quasi-periodic oscillations (QPOs) with frequencies from 18 Hz to
$\simeq $ 2 kHz in SGR 1806
\citep{israel_etal05,ws06,sw06,hambaryan_etal11}, and 28 Hz to 155 Hz in 
SGR 1900 \citep{sw05}. In these two giant flares, the Fourier power of
the burst drops into the noise above $\sim 100$ Hz, and in the case of
SGR 1806, all of the Fourier power above $\sim 100$ Hz is in the
QPOs. Measured spin down parameters imply surface dipole fields of
$6\times 10^{14}$ G for SGR 0526-66 \citep{tiengo_etal09}, $7\times
10^{14}$ G for SGR 1900+14 \citep{mereghetti_etal06}, and $2\times
10^{15}$ G for SGR 1806-20 \citep{nakagawa_etal08}. These strong
inferred fields, along with other properties such as quiescent
brightness \citep{td95,td96,wt06}, establish these objects as
magnetars.

Anomalous x-ray pulsars (AXPs) are also magnetars that exhibit
bursts that are in most respects like SGR flares, but with a wider
range of burst durations. These bursts are generally less energetic
than SGR flares, with the most energetic bursts showing much harder
spectra than are seen in SGR bursts (see, \eg,
\citealt{gavriil_etal04} and \citealt{kaspi07}). 

In addition to these high-field objects, there are now three known
``low-field'' magnetars that produce small flares. SGR 0418+5729 has
an inferred dipole field of $6\times 10^{12}$ \citep{rea_etal13a},
typical of radio pulsars. Swift J1822.3-1606 has an inferred dipole
field of $\sim 2\times 10^{13}$ G \citep{rea_etal12,scholz_etal12},
while 3XMM J185246.6+003317 has an inferred field of $<4\times
10^{13}$ G \citep{rea_etal13}. These sources show that magnetar
activity does not require high dipolar fields. Instead, the activity
could be driven by decay of multipolar components that could be
an order of magnitude or more larger than the dipolar component
\citep{braithwaite09}. These bursts are rather different than SGR
bursts. The total energy release is $\lap 10^{41}$ erg, like small
flares in SGRs, but with decay times of hundreds of days. The timing
resolution is insufficient to ascertain how the rise times compare to
bursts in SGRs. The long relaxation time is consistent with the
thermal relaxation time predicted for the crust
\citep{bc09,scholz_etal12}.

While energetics considerations strongly suggest that SGR flares 
are driven by the release of magnetic energy (see \S
\ref{energetics}), the trigger
mechanism for flares remains unknown. \citet{td01} have argued that
the elastic crust cannot store nearly enough elastic energy to power a
giant flare, but that the crust can act as a gate that holds back much
more magnetic energy. In this connection, \citet{td01} have proposed
that the core field evolves into a twisted configuration through Hall
drift, stressing the crust until it suddenly yields, producing
impulsive energy release throughout the stellar interior.  A second
possibility is that a magnetohydrodynamic (MHD) instability occurs in
the liquid core
\citep{td95}. This possibility might appear unlikely, since the core is
expected to be superconducting and the proton-electron mixture will
have a very high electrical conductivity, but as \citet{fr77} pointed
out and
\citet{td95} have stressed, if the core
field is stabilized by crust currents, and these currents have time to
decay, the core field could suffer an interchange
instability over MHD
time-scales. Moreover, if the core field exceeds $\sim 10^{16}$ G,
the core will not be superconducting, the electrical conductivity will
be lower, and the field might evolve to the point that an MHD
instability occurs.  A third possibility is that the energy is released
not inside the star, but in the magnetosphere through an MHD
instability
\citep{lyutikov03,lyutikov06,komissarov_etal07,gh10}, such as the ``tearing
mode'', producing a magnetospheric explosion akin to coronal mass
ejections seen in the Sun. 

\citet{duncan98} predicted that magnetic stresses that arise as the
internal field of a magnetar evolves will eventually shear the crust to
failure, producing a flare and exciting torsional modes in the
crust.\footnote{In general excitation of many stellar modes,
including $p$-modes, $g$-modes, and $f$ modes should occur, but
torsional modes have the lowest frequencies and would be the easiest
to detect. } The QPOs
seen in the giant flares of SGRs 1900 and 1806 were interpreted
initially as crustal modes (\eg,
\citealt{piro05,sa06,wr07,lee07,sotani_etal07,sw09}). Subsequent work
has accounted for magnetic coupling between the crust and liquid core,
and attributes QPOs to {\em global} magneto-elastic oscillations of
the neutron star (\eg,
\citealt{levin06,gsa06,levin07,sotani_etal08,cbk09,csf09,ck11,vl11,gabler_etal11,vl12,gabler_etal12,gabler_etal13,pl13a,pl13b}).
A crucial ingredient in the interpretation of QPOs as stellar
oscillations is to understand how crust movement can produce the large
observed modulations of the x-ray emission by
10-20\%. \citet{timokhin_etal08} propose that twisting of the crust,
associated with a stellar mode, modulates the charge density in the
magnetosphere, creating variations in the optical depth for resonant
Compton scattering of the hard x-ray photons that accompany the
flare. In this model, the shear amplitude at the stellar surface must
be as large as 1\% of the stellar radius, and it is unknown if the
stellar crust can sustain the associated strain without failing.
\citet{dw12} have shown that beaming effects can increase the
amplitude of the QPO emission by a factor of typically several. As the
theory of neutron star ``seismology'' is further developed, the exciting
possibility of constraining the properties of dense matter and the
magnetic field configuration of the core is becoming feasible. 

This paper is concerned with the basic question of whether the energy
that drives SGR flares is stored in the stellar interior or in the
magnetosphere just before the flare occurs. A key physical feature is the
existence of a large impedance mismatch between the stellar interior
and the magnetosphere for the propagation of shear waves, as
originally pointed out by \citet{blaes_etal89}; the mismatch is due to
the fact that the wave propagation speed in the core is $\sim 10^{-3}$
of that in the magnetosphere. With minimal assumptions, I show that
shear waves produced in the core through sudden global relaxation of
the magnetic field are prevented from quickly entering the
magnetosphere by the impedance mismatch; rather, the outer regions of
the star and the magnetosphere are highly reflective to shear waves,
causing waves to remain trapped in the core for at least seconds to
minutes, and perhaps as long as minutes to hours for a realistic
magnetic field geometry. The trapping time greatly exceeds typical
rise times of $<10$ ms, requiring that the energy that powers 
both small and giant flares is stored in the magnetosphere just before the
flare. The energy could then be quickly released through a magnetic
instability as proposed by \citet{lyutikov03}; see, also, 
\cite{lyutikov06} and \cite{komissarov_etal07}. The energy could be
stored in the magnetosphere if, for example, the internal field
gradually untwists, slowly twisting the magnetosphere until it becomes
unstable \citep{lyutikov03}.

\cite{td95} and \cite{td01} have proposed that flares arise from the
deposition of magnetic energy {\em inside} the star. 
In the model of \citet{td01}, both small and giant flares are
driven by {\em sudden} relaxation of a globally-twisted internal magnetic
field, with the energy release gated by crust rigidity. When the crust
is stressed to failure in this model, the magnetic foot points are
suddenly sheared, and energy flows from the stellar interior into the
magnetosphere, producing a radiative event through the dissipation of
Alfv\'en waves and magnetic reconnection. \cite{td01} assume that the
failure of the crust along the fault allows flow of energy from the
core into the magnetosphere over a time-scale of order the Alfv\'en
crossing time of the star or shorter. As shown in this paper, though,
the impedance mismatch between the core and the magnetosphere slows
the flow of mechanical energy 
to at least $10^2$ to $10^4$ Alfv\'en crossing times (seconds to
minutes). Crust failure cannot remove this fundamental mismatch, and
so the proposal of
\citet{td95} and \cite{td01} that both large and small magnetar flares
are driven by the sudden release of magnetic energy stored {\em
inside} the star appears to be unviable.

If flares indeed originate as magnetospheric explosions, energy will be
trapped on closed magnetic field lines in the form of relativistic
Alfv\'en waves. The impedance mismatch between the magnetosphere and
the stellar interior makes the star highly reflective to these
waves. I obtain a lower limit for the time-scale required for
relativistic Alfv\'en waves excited by a magnetospheric explosion to
excite magneto-elastic modes in the star, and find that such modes
could be excited well before a giant flare ends. Hence, a viable
explanation for the QPOs is that they represent stellar oscillations
excited {\em by the magnetosphere}, not the stellar interior, provided
the excitation can occur before the waves are damped. 

In \S \ref{energetics}, I discuss general considerations of the
release of energy in giant flares. 
In \S \ref{transmission}, I formulate the problem of
transmission from the deep stellar interior to the surface in a planar
geometry, and estimate the transmission coefficient at low
frequency. In \S \ref{transmission1}, I calculate the transmission
coefficient as a function of frequency, accounting for the material
properties of the crust and the strong gradient in the wave
propagation speed. In \S
\ref{numerical_results}, I give numerical results. In \S
\ref{trapping_in_core}, I discuss trapping of energy in the core. In
\S \ref{realisticB}, I discuss how a realistic magnetic field geometry
will greatly decrease the transmission efficiency. 
In \S \ref{trapping_in_magnetosphere}, I discuss the trapping of
energy in the magnetosphere. An explanation of the similarities
and differences between flares in SGRs, low-field magnetars, and AXPs
is beyond the scope of this paper, but the basic ideas set forth here
apply to all three classes of objects.

\section{Energy Release in Giant Flares}

\label{energetics}

\subsection{Length Scale}

\label{l}

Suppose that the magnetic configuration within a volume $l^3$ inside
the star
adjusts, lowering the magnetic energy, ultimately driving a flare of 
radiative energy $E$. By
energy conservation, $E$ is bounded by 
\begin{equation}
E<\frac{B^2}{8\pi}l^3, 
\end{equation}
where $B$ is the average field strength. 
This is an upper limit since the field will not be reduced to zero, 
and the conversion of magnetic energy to radiation will not be perfectly
efficient. The length scale $l$ has a lower limit 
of 
\begin{equation}
l\gap 6\, \left(\frac{E}{10^{46}\mbox{ erg}}\right)^{1/3} 
B_{15}^{-2/3}\mbox{ km}, 
\end{equation}
comparable to the stellar radius for a giant flare. 
Here $B_{15}\equiv B/10^{15}$, and $B$ is measured in Gauss. 

\subsection{Power Spectrum of the Energy Deposition}

\label{ps}

Readjustment of the magnetic field configuration occurs through the
production of Alfv\'en waves in the core and magneto-elastic shear
waves in the crust; see \S \ref{transmission}. I will refer to both
kinds of waves as ``shear waves'', since their properties are the
same. Independent of how the magnetic energy is gated or released, the
volume $l^3$ will fill with shear waves over a time-scale $T=l/c_s$,
where $c_s$ is the speed of shear waves, typically $3\times
10^{-3}c$ throughout most of the star; see eqs. [\ref{cvc}] and
[\ref{ca}] below. The power spectrum of the shear waves that are
produced will have a cut-off at $\nu_c\sim 1/T$:
\begin{equation}
\nu_c\equiv\frac{1}{T}=\frac{c_s}{l}\lap
200\mbox{ Hz } B_{15}^{7/6} \left(\frac{E}{10^{46\mbox{
erg}}}\right)^{-1/3}, 
\label{nuc}
\end{equation}
assuming that the protons of the core form a type II superconductor,
so that $c_s$ is given by eq. [\ref{cvc}]. If the core protons are
normal, the estimate remains the same, though the scaling with $B$
changes to $B^{5/3}$. From the energy yields of the three giant flares
to date, eq. [\ref{nuc}] gives $\nu_c\sim 1.3$ kHz for SGR 1900, 500
Hz for SGR 0526, and 130 Hz for SGR 1806. 

In the picture model of \cite{td01} involving relaxation of a twisted
internal field, the relevant length
scale for the initial deposition of shear waves is the stellar radius,
and the high-frequency cut-off will be $\nu_c\sim c_s/R\sim 100$
Hz. This estimate does not depend on whether the magnetic energy is
released through a global instability, or if it is gated by crust
rigidity. 

Based on these estimates, the subsequent analysis will be for the 
propagation of seismic energy at frequencies below 1 kHz. 

\section{Transmission of Energy from the Stellar Interior to the
Magnetosphere}

\label{transmission}

I now calculate the efficiency with which energy deposited in the
stellar interior through global readjustment of the field is
transmitted to the magnetosphere.  For magnetic energy that is
released in the stellar interior, the energy will be deposited as
heat, sound waves, and shear waves. Thermal energy diffuses through
the star relatively slowly - \eg, over a time-scale of months
through the crust \citep{bc09,scholz_etal12}; the energy propagates
much more quickly to the surface as mechanical waves. The core
supports magnetic shear waves and sound waves. Shear waves propagate
along field lines, and the fluid is essentially incompressible
\citep{levin06}.

The crust supports shear waves, modified by the magnetic field, and
sound waves. If a medium that supports shear is driven to failure,
most of the wave energy will be in the form of shear waves if
the shear wave speed is less than the sound speed
\citep{blaes_etal89}.\footnote{For material failure through local
stresses, the lowest-order emission process of waves is
quadrupolar. The energy density in a wave of propagation speed $v_p$
scales as $v_p^{-6}$.} This is the case throughout the core and near
the base of the inner crust.

For energy to leave the core, it must propagate along field lines that
pass from the core, through the crust, and into the magnetosphere. The
plasma density is low in the magnetosphere, so energy propagates there
as relativistic Alfv\'en waves and as magnetosonic waves. Hence, the
most efficient way for energy to leave the core is to
propagate along field lines that point nearly radially outward. Outgoing
shear waves will couple most directly to Alfv\'en waves in the
magnetosphere, and I ignore the weaker coupling to magnetosonic
waves. I also note that each reflection of a shear wave in the star
will convert some of the energy to compressional waves. I ignore this
effect as well, and consider the problem of the propagation of shear
waves from the core and crust, and their emission from the star as
Alfv\'en waves.

Consider a simple planar geometry, with the crust-core boundary in the
$x-y$ plane at $z=0$, and the stellar surface at $z=z_s$. The magnetic
field is constant and directed along the $z$ axis. A
linearly-polarized shear perturbation of displacement $u(z,t)=u(z){\rm
e}^{-i\omega t}$ obeys \citep{blaes_etal89}
\begin{equation}
\frac{d}{dz}\left(\tilde{\mu}\frac{du}{dz}\right)
+\tilde{\rho}\,\omega^2\,u=0,
\label{wave_equation}
\end{equation}
where 
\begin{equation}
\tilde{\mu}\equiv\mu+\frac{B^2}{4\pi}.
\label{tildemu}
\end{equation}
Here $\mu$ is the material shear modulus, which is non-zero only in the
crust, and 
\begin{equation}
\tilde{\rho}\equiv\rho_d+\frac{B^2}{4\pi c^2},
\label{tilderho}
\end{equation}
where $\rho_d$ is the {\em dynamical} mass density, that is, the
mass density associated with matter that moves in response to a passing
shear wave. In the inner crust, Bragg scattering of free neutrons with the
nuclear lattice gives $\rho_d<\rho$
(\citealt{chamel05,chamel12,chamel13}; see eq. \ref{rhod}
below). The second term in eq. [\ref{tilderho}] is the contribution
of the magnetic energy to the effective mass of the matter, and 
is important only near the stellar surface, for $\rho\lap 10^8\,
B_{15}^2$ \gcc. The speed of shear waves is
$c_s=\sqrt{\tilde{\mu}/\tilde{\rho}}$. From eq. [\ref{wave_equation}],
continuity in $u$ requires that the traction $\tilde{\mu}du/dz$ be
everywhere continuous.

The protons of the outer core are expected to form a type-II
superconductor (SC) if the field is below the upper critical field
$H_{c2}\sim 10^{16}$ G. Type-II superconductivity modifies the
magnetic stress. Repeating the derivation of \citet{blaes_etal89}
using the magnetic stress tensor of \citet{ep77} gives
\begin{equation}
\tilde{\mu}=\frac{H_{c1}B}{4\pi} \qquad \mbox{(SC core protons)}, 
\end{equation}
where $H_{c1}\simeq 10^{15}$ G is the lower critical field. 
In the core, the protons and neutrons are expected to form distinct
superfluids, with negligible nuclear entrainment of the neutron mass
current by the proton mass current \citep{ch06,link12b}. Here magnetic
waves are supported by the charged component, and $\rho_d$ is nearly
equal to the proton mass density.  If the core is a type II
superconductor as predicted, magnetic disturbances propagate as {\em
vortex-cyclotron waves} ( \citealt{mendell98}) at speed
\begin{equation}
c_{vc}=\sqrt{\frac{BH_{c1}}{4\pi\rho x_p}}=
3\times 10^{-3}\,H_{c1,15}^{1/2}B_{15}^{1/2}
\left(\frac{\rho_{14}}{2}\right)^{-1/2}
\left(\frac{x_p}{0.05}\right)^{-1/2}\, c
\qquad \mbox{(SC core protons),}
\label{cvc}
\end{equation}
where $\rho_{14}\equiv 10^{14}\rho$, $\rho$ is in \gcc, and $x_p$ is
the proton mass fraction. Here fiducial values typical of the outer core have
been chosen.

If
the core protons are instead normal, waves propagate as Alfv\'en
waves at speed 
\begin{equation}
c_{A}={\frac{B}{\sqrt{4\pi\rho x_p}}}=
3\times 10^{-3}\, B_{15}
\left(\frac{\rho_{14}}{2}\right)^{-1/2}
\left(\frac{x_p}{0.05}\right)^{-1/2}\, c
\qquad \mbox{(normal core protons)}
\label{ca}
\end{equation}

In the magnetosphere, $\tilde{\mu}=B^2/4\pi$ and $\tilde{\rho}\simeq
B^2/4\pi c^2$, and the Alfv\'en waves are relativistic:
\begin{equation}
\frac{d^2u}{dz^2}+\frac{\omega^2}{c^2}u=0.
\end{equation}
where the wavenumber in the magnetosphere is $k=\omega/c$. 

The wavenumber in the core is $k_c=\omega_s/c_s$, where $c_s=c_{vc}$
for SC core protons and $c_s=c_A$ for normal core protons.  For
sufficiently low frequencies that $k_c\Delta R<<1$ is satisfied,
corresponding to $\nu\lap 100$ Hz, the wave is insensitive to the
gradients in $\tilde{\mu}$ in the crust, and the crust can be treated
as a thin discontinuity; crust structure is unimportant in this
limit. In this case, the energy transmission coefficient takes the
familiar form
\begin{equation}
T=\frac{4(\tilde{\mu}_ck_c)(\tilde{\mu}_m k)}{(\tilde{\mu}_m k
+\tilde{\mu}_ck_c)^2}, 
\label{Tlowf}
\end{equation}
where $\tilde{\mu}_c=BH_{c1}/4\pi$ for a superconducting core, $B^2/4\pi$
for a core of normal protons and superfluid neutrons, and
$\tilde{\mu}_m=B^2/4\pi$ for the magnetosphere. (Recall that $B$ is
constant in the assumed planar geometry).  Typically
$\tilde{\mu}_ck_c>>\tilde{\mu}_mk$, giving, for a superconducting core, 
\begin{equation}
T\simeq 4\left(\frac{B}{H_{c1}}\right)\frac{c_{vc}}{c}
\simeq 10^{-2}\, \left(\frac{B}{H_{c1}}\right)^{3/2}
\left(\frac{x_p}{0.05}\right)^{-1/2}
\left(\frac{\rho_{14}}{2}\right)^{-1/2}
\qquad
\mbox{(SC core protons)}.
\label{Tsc}
\end{equation}
while for a core of normal protons 
\begin{equation}
T\simeq 4\frac{c_A}{c}
\simeq 10^{-2}\, B_{15}
\left(\frac{x_p}{0.05}\right)^{-1/2}
\left(\frac{\rho_{14}}{2}\right)^{-1/2}
\qquad
\mbox{(normal core protons)}.
\label{Tn}
\end{equation}
Because $\tilde{\mu}_ck_c>>\tilde{\mu}_mk$, there is a strong
impedance mismatch between the core and the magnetosphere, giving
$T<<1$ for $\nu\lap 100$ Hz. 
We will see that $T$ is further reduced by the structure of the crust
for 
$\nu\gap 100 $ Hz. 

Energy that is released primarily in the core cannot 
propagate directly to the surface, but becomes trapped in the core. 
For energy to propagate into the
magnetosphere, it must then propagate from the core and through the
crust, suffering multiple reflections before escaping to the
magnetosphere. 

\section{Energy Transmission Coefficient}

\label{transmission1}

I now turn to an exact calculation of the energy transmission coefficient
$T$ for $\nu\gap 100$ Hz, when $k_c\Delta R <<1$ is not satisfied,
and crust structure has important effects for wave propagation. 

The shear modulus in the crust, ignoring magnetic effects, is
\citep{strohmayer_etal91} 
\begin{equation}
\mu=\frac{0.1194}{1+0.595(173/\Gamma)^2}\frac{n_i(Ze)^2}{a}, 
\end{equation}
where $n_i$ is the number density of ions of charge $Ze$, $a$ is the
Wigner-Seitz cell radius given by $n_i4\pi a^3/3=1$, and $\Gamma\equiv
(Ze)^2/(akT)$ where $k$ is Boltzmann's constant. Typically in the
crust, $\Gamma>>173$ and the second term in the denominator is
negligible. For the composition of the crust, I use the results of
\citet{hp94} for the outer crust, and the results of
\citet{dh01} for the inner crust, conveniently expressed analytically
by \citet{hp04}, who treat densities from $10^5$ \gcc\ to above
nuclear density. The treatment by \cite{dh01} of the inner crust gives
somewhat higher values of the shear modulus at the base of the crust
than do other studies. The equation of state
of \cite{akmal_etal98}, for example, gives a shear speed at the base of the
crust that is about 0.6 the shear speed 
of \citet{dh01}, and a corresponding shear modulus that is smaller by
a factor of about 2.8. A higher shear speed in the crust decreases the
impedance mismatch with respect to magnetosphere, giving somewhat
higher values for the transmission coefficient that most other choices of
the shear modulus would give. 

\begin{figure*}
\centering
\includegraphics[width=.6\linewidth]{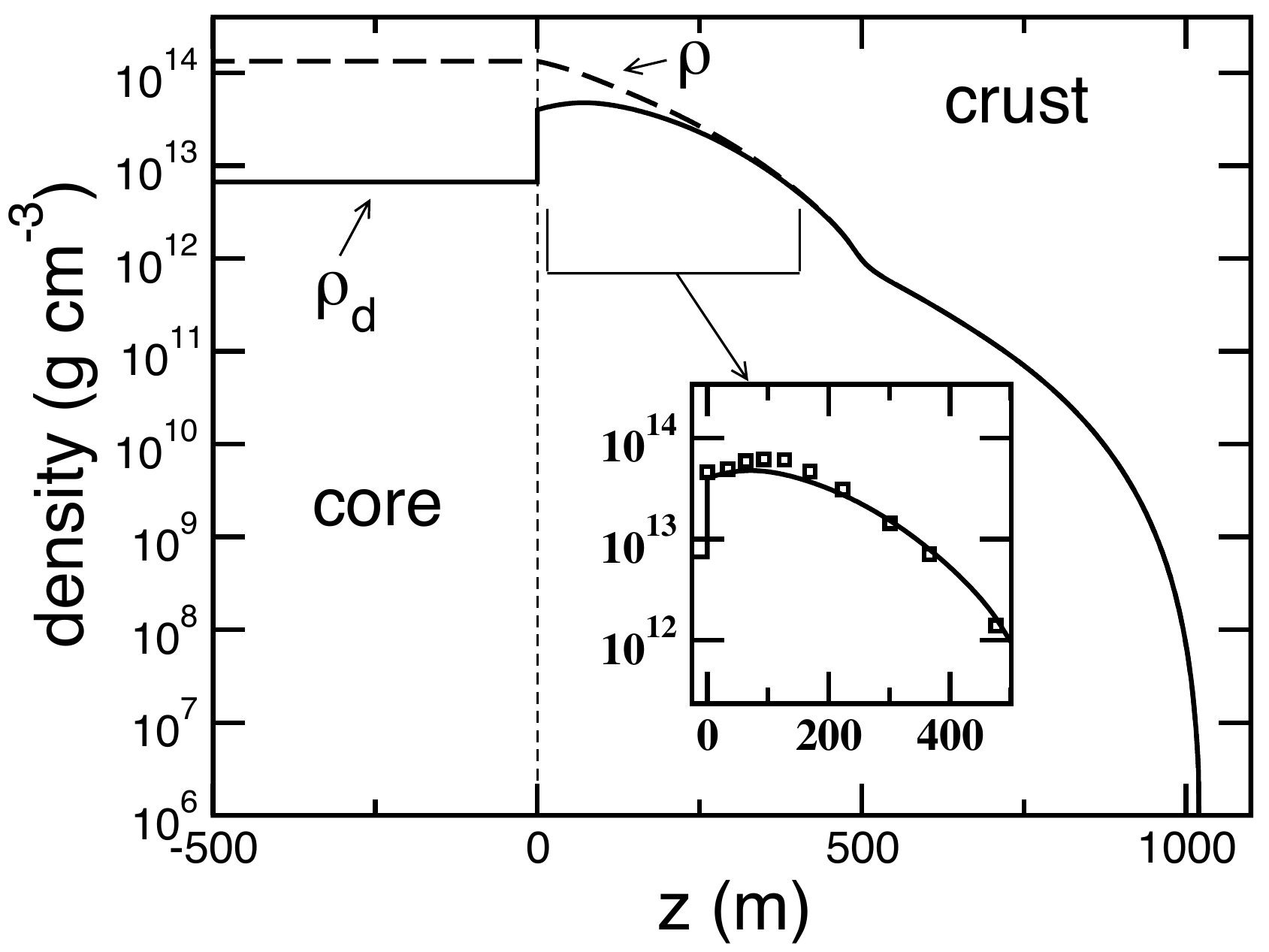} 
\caption{The density profile in the inner crust and outer core. The
dashed curve gives the baryonic mass density profile from the solution
to eq. [\ref{he}] with the equation of state of \citet{hp04}. The
solid curve shows the dynamical mass density of
eq. [\ref{rhod}]. In the denser regions of the inner crust,
$\rho_d<\rho$ from the effects of nuclear entrainment
\citep{chamel05,chamel12,chamel13}; see eq. [\ref{rhod}]. The inset
shows detail of the region where nuclear entrainment effects are most
important; the squares are from the calculations of
\citet{chamel12}. In the core,
$\rho_d$ is equal to the proton mass density, here fixed to be the
density $x_p\rho_c=6.5\times 10^{13}$ \gcc\ at the crust-core interface.}
\label{density}
\end{figure*}

For a barytropic equation of state $p(\rho)$, the density profile in the crust,
neglecting the effects of General Relativity, follows from the equation
of hydrostatic equilibrium 
\begin{equation}
\frac{d\rho}{dr}=-g\rho(r)\left(\frac{dp}{d\rho}\right)^{-1},
\label{he}
\end{equation}
where $g$ is the gravitational acceleration. Henceforth I fix
$g=2\times 10^{14}$ cm s$^{-2}$, appropriate to a neutron star of 1.4
solar masses and a radius of 10 km.  I take the
crust to dissolve into the core at $\rho_c=1.3\times 10^{14}$ \gcc,
about half of nuclear saturation density. Under these assumptions, the
crust thickness $\Delta R$ is almost exactly 1 km. The density
profile in the crust is given by the dashed line in
Fig. \ref{density}.

I take the core to be of constant density and infinite extent 
for $z<0$, with a wave incident on the crust-core interface,
and a reflected wave:
\begin{equation}
u(z)=A{\rm e}^{ik_cz}+B{\rm e}^{-ik_cz}, 
\end{equation}
where $A$ and $B$ are constants. 
Requiring continuity in $u$ and $\tilde{\mu}\,du/dz$ at the crust-core
interface ($z=0$), gives
the transmission coefficient
\begin{equation}
T=1-\left|\frac{B}{A}\right|^2=
1-
\left|
\frac{\tilde{\mu}(0-)u(0+)k_c+i\tilde{\mu}(0+)u^\prime(0+)}
{\tilde{\mu}(0-)u(0+)k_c-i\tilde{\mu}(0+)u^\prime(0+)}
\right|^2, 
\label{T}
\end{equation}
where a prime denotes a derivative in $z$. 

At the surface $z=z_s=\Delta R$, there is only an outgoing
relativistic Alfv\'en wave ${\rm e}^{ik(z-z_s)}$ with
$k=\omega/c$. Since $A$ and $B$ are unspecified, 
$u$ is conveniently fixed to unity at the surface. Continuity in
$\tilde{\mu}\,du/dz$ gives the
surface boundary condition
\begin{equation}
\tilde{\mu}(z-z_s=0-)u^\prime (z-z_s=0-)=ik\tilde{\mu}(z-z_s=0+)
\quad
\mbox{with}
\quad
u(z-z_s=0-)=1.
\end{equation}
The amplitude $u$ becomes complex for $z<z_s$. Calculation of the 
quantities $u(0+)$ and $u^\prime(0+)$ by numerical
integration from the surface to $z=0$ gives the transmission
coefficient from eq. [\ref{T}].

\begin{figure*}
\centering
\includegraphics[width=.6\linewidth]{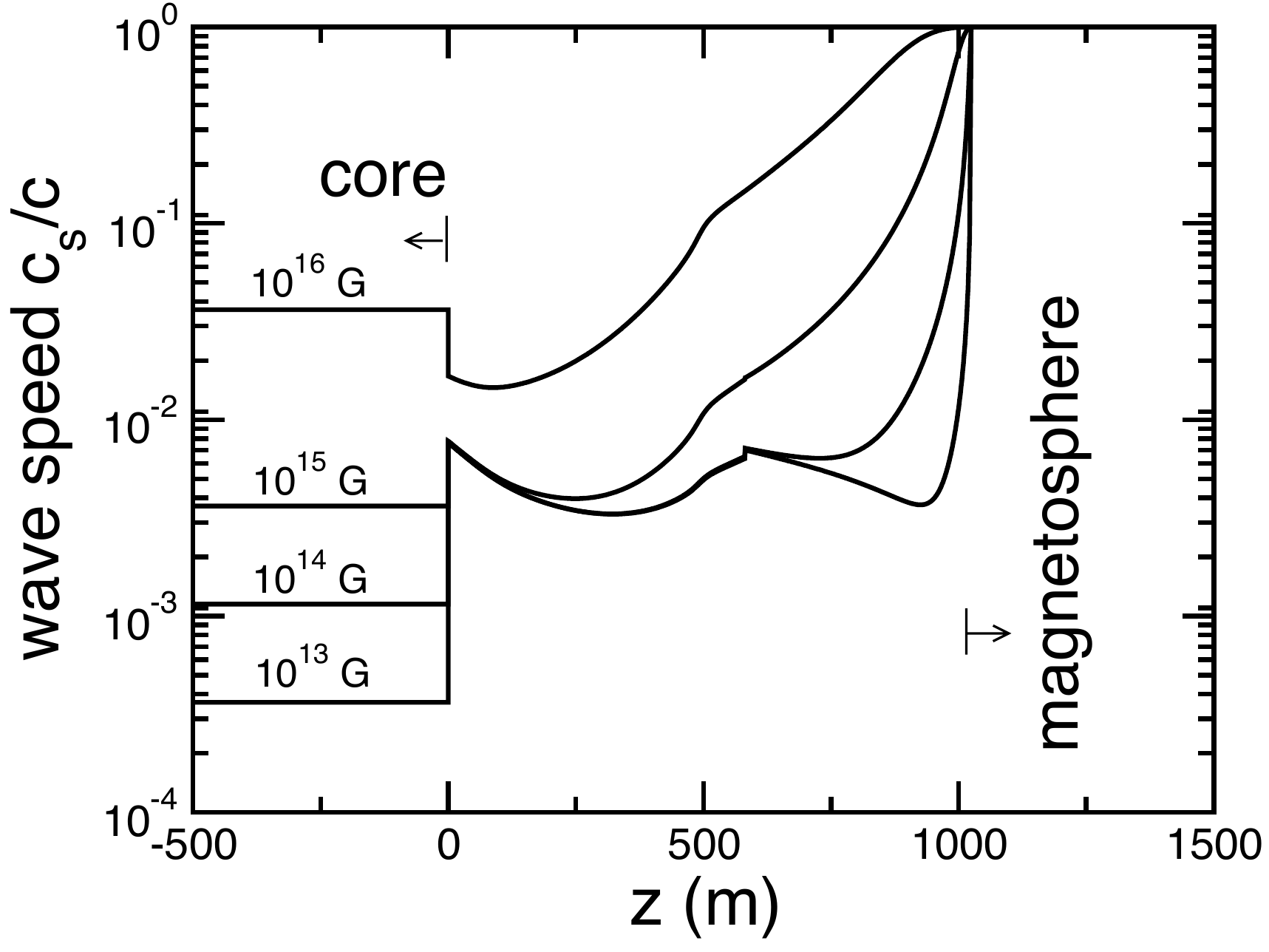} 
\caption{The shear speed $c_s=\sqrt{\tilde{\mu}/\tilde{\rho}}$ as a function of position. As the magnetic field is
increased, the effective shear modulus $\tilde{\mu}$ increases in the
crust; see eq. [\ref{tildemu}]. For $B=10^{16}\mbox{ G}\simeq H_{c2}$,
the core is
taken to be normal. The core neutrons are assumed to be superfluid. 
Outside the star, the shear speed is $c$, the
speed of a relativistic Alfv\'en wave. There is a small jump in $\mu$
at $z\simeq 600$ m, corresponding to the neutron drip density that is
washed out by magnetic stresses for $B\gap 3\times 10^{14}$ G. }
\label{cs}
\end{figure*}

In the inner crust, most of the baryonic mass is in the form of
superfluid neutrons. As a shear wave passes, a fraction of the
superfluid neutrons is non-dissipatively entrained by the nuclear
clusters through Bragg scattering \citep{chamel05,chamel12}; the
remaining ``conduction'' neutrons, that is, the neutrons that are not
entrained,  must be subtracted from the baryonic
density to give the appropriate dynamical density \citep{pcr10}:
\begin{equation}
\rho_d=\rho\,(1-n_n^c/\bar{n})\equiv f\rho,
\label{rhod}
\end{equation}
where $n_n^c$ is the number density of conduction neutrons and
$\bar{n}$ is the average baryon density in a unit cell. The speed of
shear waves in the crust (for $B=0$) is $c_s=\sqrt{\mu/\rho_d}$; the
existence of conduction neutrons increases the shear speed. To account
for this effect, a fitting formula for $f$ that gives a good
approximation to the results of \citet{chamel12} is 
\begin{equation}
f=1-\frac{8.8\bar{n}}{1+{\rm exp}\left(10^3(\bar{n}-0.09)\right)}
\qquad\qquad \bar{n}<0.08\mbox{ fm$^{-3}$}.
\end{equation}
$f$ stays near
unity up to $\bar{n}\simeq 0.04$ fm$^{-3}$, before falling to $\sim
0.35$ at $\bar{n}=0.08$ fm$^{-3}$, at which point the inner crust
ends. The inset of Fig. \ref{density} shows the assumed
dynamical density based on the results of \citet{chamel12}. 
The shear speed in the star is shown in
Fig. \ref{cs}.

The inner core could reach a density of $5-10\rho_c$, but the
gradients in $\tilde\rho$ and $\tilde\mu$ are always much less than in
the inner crust. Some of the energy that is 
propagating outward will be reflected by the relatively small gradients
in $\tilde{\rho}$ and $\tilde{\mu}$ in the core. Treating the core as
having constant density  
slightly overestimates the transmission coefficient. Also, 
the choice of a constant field in the $z$ direction is the most
favourable geometry for the propagation of energy out of the core,
through the crust, and into the magnetosphere. Any other field
configuration will lead to more effective reflection of energy from
the crust and the magnetosphere back into the core. For realistic
field configurations, this effect could be large. As argued in
\S \ref{realisticB}, these calculations of
the transmission coefficient probably represent {\em significant}
overestimates, so the energy transmission efficiency calculated in this
paper is a robust upper limit. 

\section{Numerical Results}

\label{numerical_results}

\begin{figure*}
\centering
\includegraphics[width=.6\linewidth]{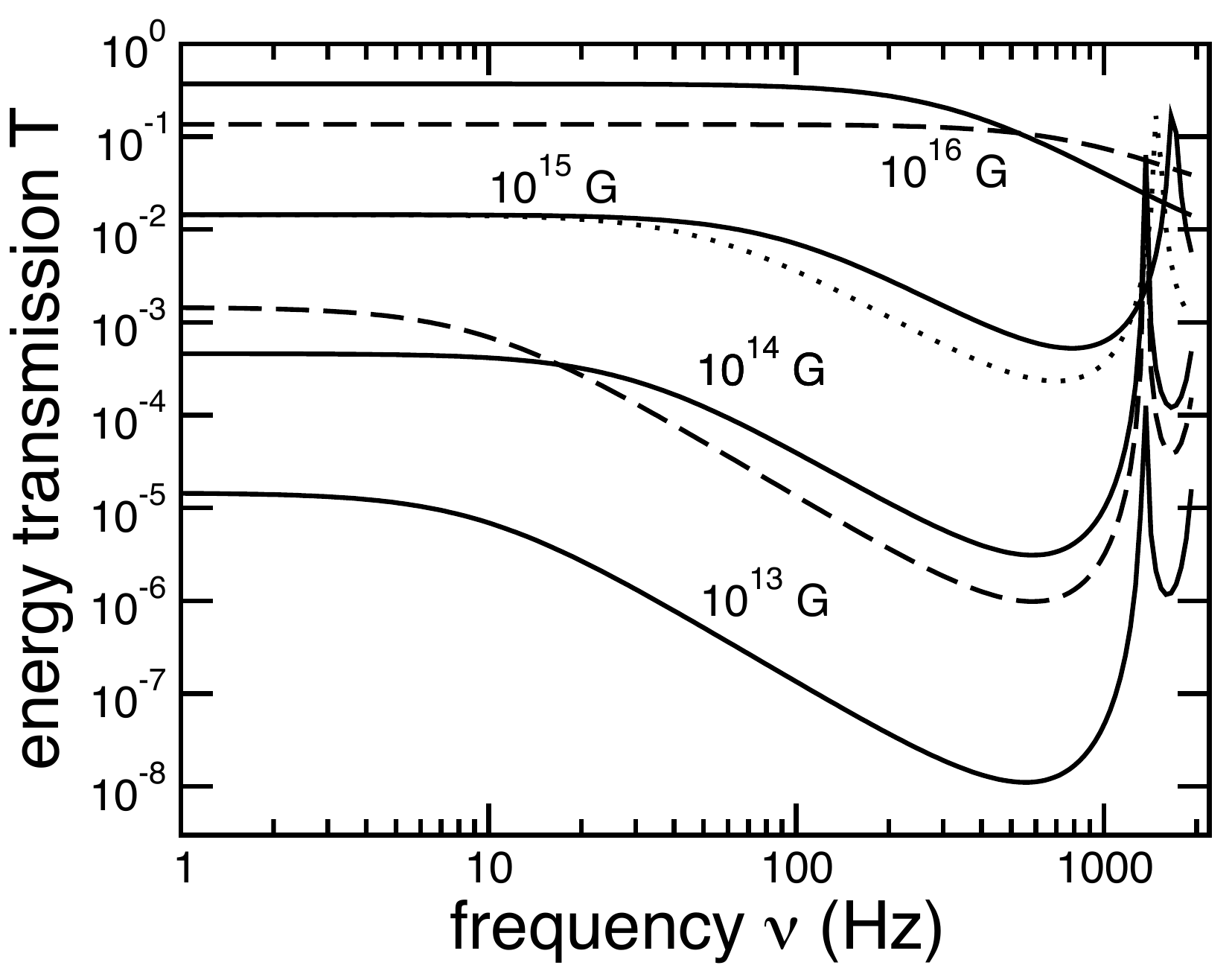} 
\caption{The transmission coefficient as a function of frequency. For
frequencies above $10-100$ Hz (depending on $B$), gradients in
$\tilde{\mu}$ and $\tilde{\rho}$ increase the reflection of the wave
back into the core, reducing $T$. Above $\sim 1$ kHz, transmission
resonances appear in the simplified planar geometry. The solid curves
show $T$ for a core of superconducting protons, and the dashed curves
are for normal protons; the curves are the same for
$B=H_{c1}=10^{15}$ G. The dotted curve for $B=10^{15}$ shows the
effect that perfect nuclear entrainment in the inner crust would have,
corresponding to $\rho_d=\rho$. The core neutrons are assumed to be
superfluid.  }
\label{Tplot}
\end{figure*}

Calculations of the transmission coefficient are shown in Fig. \ref{Tplot}
for different values of the magnetic field strength, assuming that the
core neutrons are superfluid. For each solid
curve, the core was assumed to be a type II superconductor. For each
dashed curve, the core protons are assumed to be normal - the two
curves coincide for $B=H_{c1}=10^{15}$ G. For $\nu\lap 100$ Hz, $k_c\Delta
R<<1$, and $T$ is nearly independent of frequency. For $B<H_{c1}$, type
II superconductivity increases the magnetic stress in the core by a
factor $(H_{c1}/B)^{1/2}$ relative to the normal case (see
eqs. \ref{Tsc} and \ref{Tn}), giving a corresponding increase in $T$ by
decreasing the impedance mismatch between the core and the
magnetosphere; see eq. [\ref{Tlowf}]. For $B>H_{c1}$, the situation is
reversed, and superconductivity decreases $T$. A field of $B=10^{16}$
G is close to the upper critical field $H_{c2}$ above which
superconductivity is destroyed; only the
dashed curve is likely be relevant for $B=10^{16}$ G.

Above $\nu\sim 100$ Hz, gradients in the crust of the density and the
shear modulus act as an effective potential which partially reflects
the wave back into the core, reducing $T$. At $\nu\sim 1$ kHz, the
solution shows strong transmission resonances that are a consequence
of the assumed planar geometry and constant field. More realistic
field structure and energy deposition geometry will eliminate these
resonances.

The dotted line shows the effect of perfect entrainment by nuclear
clusters in the inner crust ($f=1$). Entrainment increases the
effective mass of nuclear clusters, reducing the shear speed for zero
field, and reducing $T$. Because a fraction of the
neutron superfluid does not move with the nuclei, the shear speed is
increased, and $T$ increases by a factor of $\sim 3$ at most. 

The propagation of seismic energy from the crust into the
magnetosphere was studied by \citet{blaes_etal89} in the context of
gamma-ray bursts from neutron stars assuming shallow energy
deposition, at a density less than the neutron drip density. They
found the existence of an evanescent wave zone very close to the
stellar surface that is not found in the analysis given here. To
evaluate the transmission coefficient, they evaluated
$\tilde{\mu_c}k_c$ in eq. [\ref{Tlowf}] at the base of the evanescent
wave zone. Eq. [\ref{Tlowf}] applies only at low frequency, and only
for the case that a wave zone exists for $z<0$. \citet{blaes_etal89}
considered transmission for frequencies in the range $10^3<\nu<10^5$
Hz. Given the different boundary conditions and frequency regimes, a
direct comparison to their work is not possible.

\section{Trapping of Energy in the Core}

\label{trapping_in_core}

\begin{figure*}
\centering
\includegraphics[width=.6\linewidth]{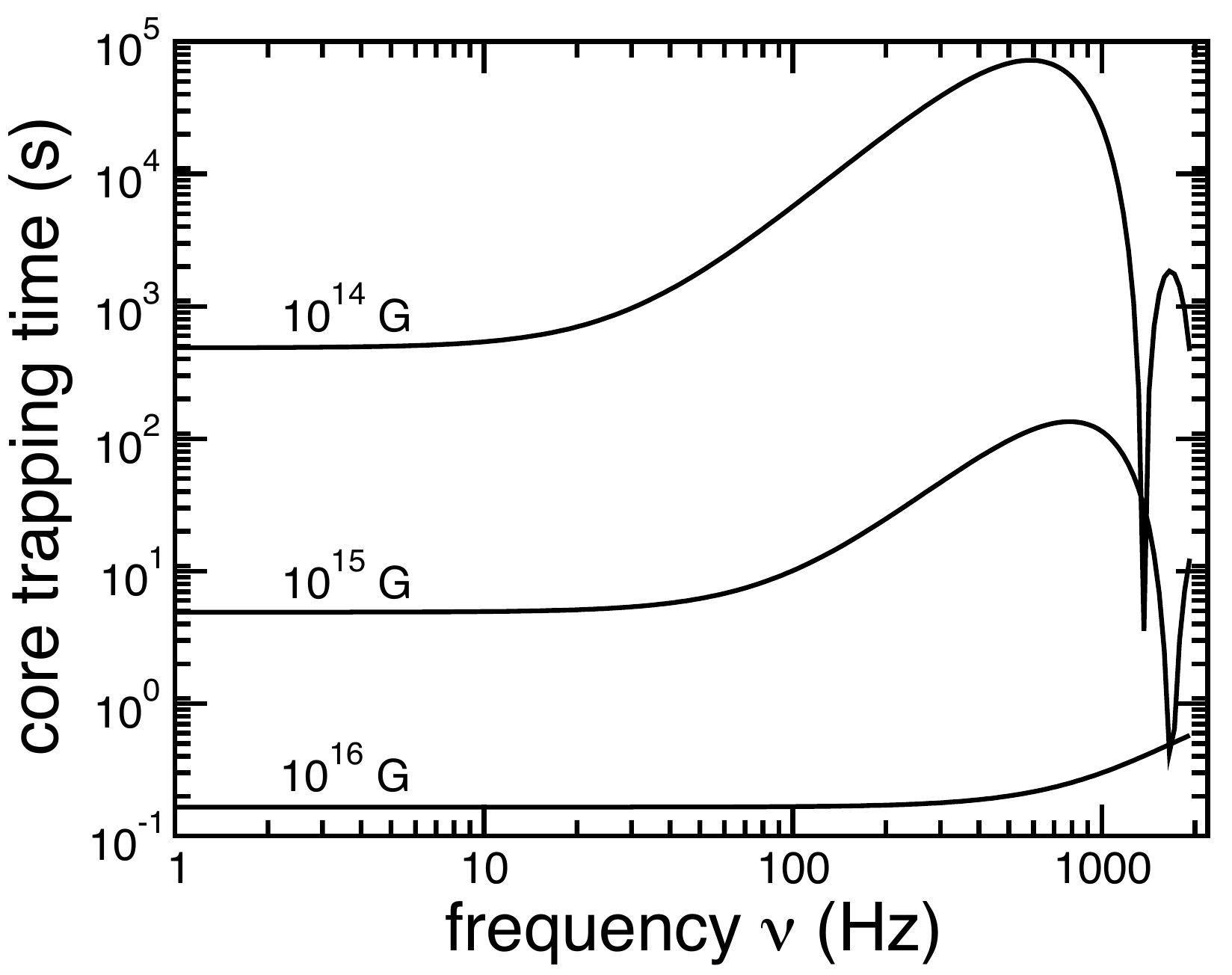} 
\caption{The trapping time for waves in the core. For $B=10^{16}$ G,
the core protons are assumed to be normal. These time-scales are lower limits
(see the text).  }
\label{ttrap}
\end{figure*}

Shear waves trapped in the core carry energy across the core at a
speed equal to the wave group velocity, $c_s=c_{cv}$ for
superconducting core protons, and $c_s=c_A$ for normal protons.  The
wave crossing time is $\simeq 2R/c_s$. The `attempt frequency' is
$c_s/2R$, with an energy transmission probability $T(\nu)$ per attempt, so the
energy transmission rate is $\sim (c_s/2R)T(\nu)$ for a mode of
frequency $\nu$.  The associated trapping time for energy in the core
is thus
\begin{equation}
t_{\rm trap}\simeq \frac{2R}{c_s T(\nu)}.
\end{equation}
The trapping time is shown in Fig. \ref{ttrap}. The spin-down rates
for both SGR 1900 and SGR 1806 imply a dipole field of strength
$B\simeq 10^{15}$ G. The trapping time is $\sim 5$ s below 100 Hz, and
up to $\sim 100$ s at higher frequencies. These time-scales greatly
exceed the observed rise time of $\lap 10$ ms that is seen in both
giant flares and in small bursts. Even if $B=10^{16}$ G, which is much
larger than the field implied by the observed spin-down rates, energy
cannot enter the magnetosphere from the core nearly quickly enough to explain
observed rise times. Note that these trapping times represent lower
limits, since the most favourable magnetic geometry for coupling of the
stellar interior to the magnetosphere was assumed. These results
strongly suggest that the flares are powered by the release of
magnetic energy directly into the magnetosphere, not in the core. 

\section{Effects of Realistic Field Geometries}

\label{realisticB}

The analysis so far treats the magnetosphere as having infinite
extent. Energy is transferred most efficiently to the magnetosphere
when it excites field lines that are long enough to resonate with the
wave that propagates from the core; coupling to shorter field lines
will be greatly reduced, as will the transmission coefficient
integrated over the stellar surface. For a
field line of length $L$ with two fixed foot-points on the star, the
fundamental frequency is $c/2L$. Field lines with fundamental
frequencies below 1 kHz are longer than 150 km. Most of the field
lines that emerge from star are  much shorter length than this, so
the mechanical coupling between the stellar interior and the
magnetosphere is poor. The energy flux into the magnetosphere will be
reduced by a factor of $A/4\pi R^2$
with respect to the spherically symmetric situation, where $A$ is the
area of the star that is connected to field lines that are long enough
to resonate with seismic waves. Though the transmission coefficient
$T(\nu)$ has been calculated in a planar approximation for simplicity,
a good final estimate for the transmission rate integrated over the
stellar surface is $T(\nu)A/4\pi R^2$. 

To estimate $A$, recall that the field line configuration from a
magnetic dipole is given by 
\begin{equation}
\frac{r}{R}=\left(\frac{\sin\theta}{\sin\theta_L}\right)^2, 
\label{dipoleline}
\end{equation}
where $(r,\theta)$ are spherical coordinates measure with respect to
the magnetic dipole moment, and $\theta_L$ is the
angle the field line takes at $r=R$ (the stellar surface in this
simple approximation). Integration of eq. [\ref{dipoleline}] along any
given line gives the length of the field line
$L$ in the limit $\theta_L<<1$:
\begin{equation}
L\simeq 3.6 R\sin^{-2}\theta_L.
\label{L}
\end{equation}
If the power spectrum of excited magnetospheric waves
ends at a cut-off $\nu_c$, corresponding to a value $\theta_{L,{\rm
max}}$ from eq. [\ref{L}], the area of the magnetic polar region
through which Alfv\'en waves enter the star is approximately $A\simeq
\pi(R\theta_{L,{\rm max}})^2\simeq 10^{-4}4\pi R^2\,\nu_c
(\mbox{Hz})$. Taking $\nu_c=1$ kHz, indicates that the energy transfer
into the magnetosphere could be reduced by a factor of about 10,
raising the curves in Fig. \ref{ttrap} by the same factor. If $\nu$ is
closer to 100 Hz, as suggested by the estimates of \S \ref{ps}
assuming global readjustment of interior magnetic field, the curves of
Fig. \ref{ttrap} go up by a factor of about $10^2$. The trapping time
for $B=10^{15}$ G becomes $\sim 400$ s to $\sim 3$ h for $\nu<1$
kHz. 

\section{Trapping of Energy in the Magnetosphere and QPO Excitation}

\label{trapping_in_magnetosphere}

That the trapping time for seismic energy in the star is much longer
than observed rise times suggests that flares are driven by the
release of energy stored in the 
magnetosphere, where magnetic energy might be converted to Alfv\'en
waves and radiation much more quickly. If QPOs represent stellar
oscillations, they might be excited by the absorption of the star of
relativistic Alfv\'en waves from the magnetosphere. I now estimate 
this time-scale.

\begin{figure*}
\centering
\includegraphics[width=.6\linewidth]{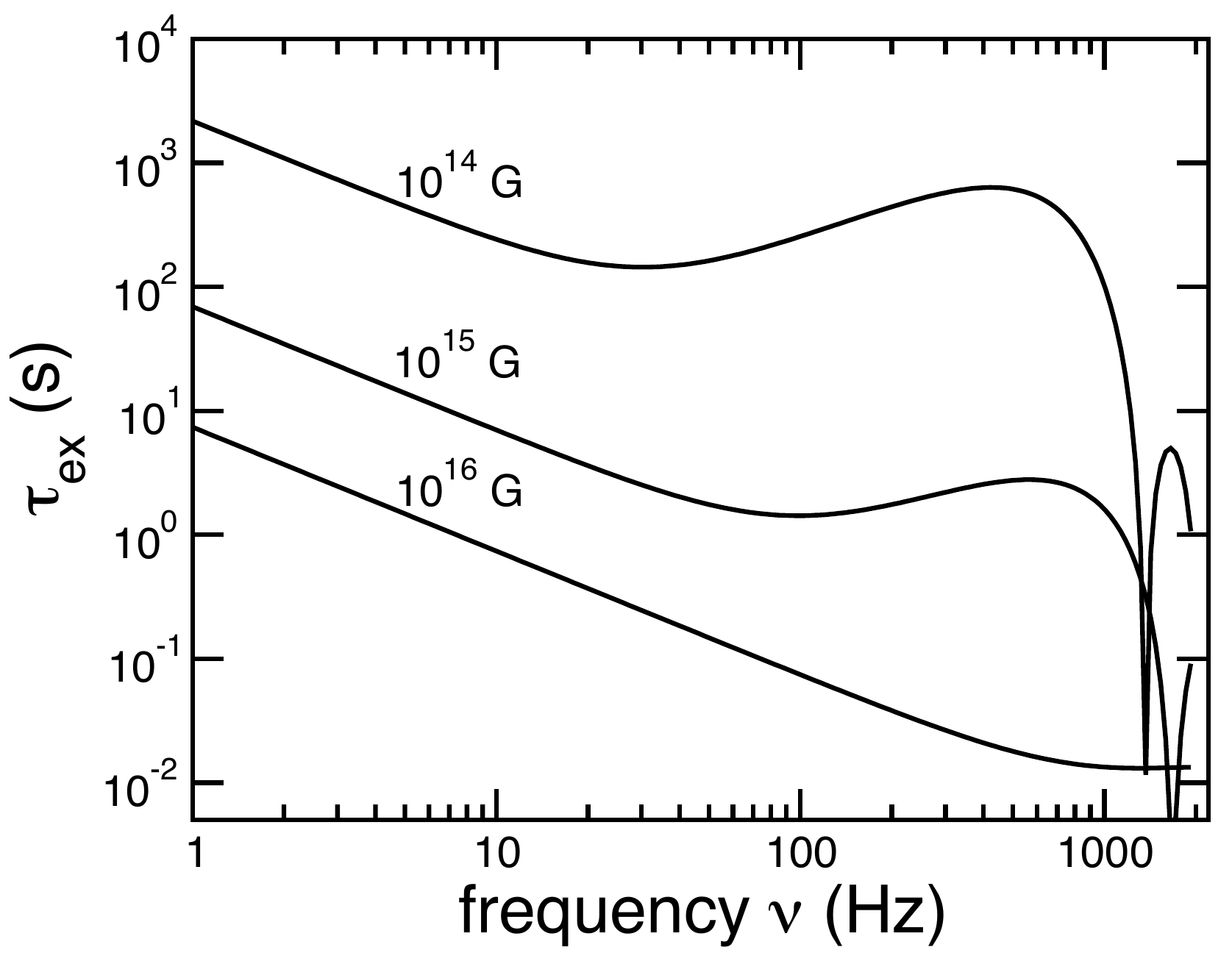} 
\caption{The time-scale over which Alfv\'en waves in the magnetosphere
can excite stellar modes. The core protons are 
assumed to be normal for $B=10^{16}$ G.}
\label{tex1}
\end{figure*}

By energy conservation, the transmission coefficient for the
excitation process is the same $T(\nu)$ calculated above. Since
$T(\nu)$ is much less than unity for $1\mbox{ Hz }<\nu<1$ kHz, the
stellar surface is highly reflective to Alfv\'en waves. As a result,
energy deposited in the magnetosphere will be trapped on closed field
lines before it is either absorbed by the star or dissipated in the
magnetosphere. A mode of frequency $\nu$ supported by closed field
lines in the magnetosphere will bounce off the stellar surface at a
rate $\nu$, with an absorption probability $T(\nu)$ at each bounce.
Ignoring dissipation in the magnetosphere, the absorption rate by the
star of a magnetospheric Alfv\'en wave of frequency $\nu$ is $\nu
T(\nu)$ for a planar geometry. As the energy is absorbed, it excites
primarily shear waves in the crust and core (see \S
\ref{transmission}).  The characteristic excitation time-scale for a
stellar mode of frequency $\nu$ by an Alfv\'en wave in the
magnetosphere of frequency $\nu$ is
\begin{equation}
\tau_{ex}\sim [\nu T(\nu)]^{-1}
\label{tex}
\end{equation}
This time-scale is shown in Fig. \ref{tex1}. We see that for
$B\ge 10^{15}$ G, most of the energy deposited in the magnetosphere could
be absorbed by the star before the end of a giant flare, and so there
could be sufficient time for relativistic Alfv\'en waves in the
magnetosphere to excite stellar modes and associated QPOs. The
strongest QPO seen in the tail of the giant flare in SGR 1806, at 92.5
Hz, was estimated to appear about two minutes into the
flare \citep{israel_etal05,sw06}. 

Eq. [\ref{tex}] is a crude estimate. The field geometry near the
surface of the star and inside the star is likely to be quite
complicated, and certainly not everywhere perpendicular to the surface
as assumed in this simple planar treatment. If the magnetospheric
structure constrains the delivery of Alfv\'en energy to only small
patches on the stellar surface, the energy transfer rate into the star
could be greatly reduced, making $\tau_{ex}$ much longer. If this is
the case, the interpretation of QPOs as magneto-stellar oscillations
could be problematic.  Calculations with more realistic field
geometries are needed to resolve this issue; 
$\tau_{ex}$ calculated here is most likely a lower limit. 

Suppose, however, that the magnetosphere changes its structure
instantaneously. Now there is no time-scale in the problem, and all
frequencies will be excited in the magnetosphere; $f$ modes and
torsional modes could be excited to large amplitudes
\citep{lv11}. More realistically, the time-scale for adjustment of the
inner magnetosphere through an instability is $\sim R/c\simeq 30$
$\mu$s, implying a cut-off frequency of $\sim 30$ kHz.  The
transmission coefficient generally increases with frequency, and high
frequency waves could enter the star relatively easily.

A big uncertainty is the damping rate of Alfv\'en waves in the
magnetosphere. In a magnetized plasma with a gradient in the Alfv\'en
velocity, the case in the neutron star magnetosphere, dephasing of
Alfv\'en waves drives currents and the wave can be quickly damped by
electrical resistivity \citep{hv83}. If the damping rate is too fast,
the magnetosphere cannot excite global magneto-elastic modes. This
problem merits further study.

\section{Conclusions}

The main conclusion of this paper is that the large impedance mismatch
between the neutron star interior and the magnetosphere causes energy
exchange between the two to be relatively slow. If the energy that
drives a flare, either a small flare or a giant flare, is driven by
{\em sudden}, global relaxation of the internal magnetic field, the
trapped seismic energy takes at least seconds to minutes to reach the
magnetosphere for $B=10^{15}$ G (see Fig. \ref{ttrap}), and
possibility as long as minutes to hours for a realistic magnetic field
geometry, in any case much longer than the observed rise times of
$<10$ ms. This conclusion rules out models of flares powered by {\em
sudden}, internal magnetic relaxation (\eg, \citealt{td95,td01}).
Crust failure cannot remove the fundamental impedance mismatch that
limits the transfer of shear-wave energy from the core into the
magnetosphere.  {\em The energy that drives a flare must be stored in
the magnetosphere}. One way this could happen is if the internal field
gradually untwists, slowly twisting the magnetosphere until it becomes
unstable \citep{lyutikov03}. A lower limit to the rise time will be
determined by the time-scale over which the instability develops. This
time-scale could be $\lap 10$ ms for the tearing mode
\citep{lyutikov06,komissarov_etal07}. The rise time of the observed
flux could be longer, depending on the emission processes that
accompany the instability.

The lower limits on the excitation time-scales of stellar modes by
relativistic Alfv\'en waves in the magnetosphere given in
Fig. \ref{tex1} indicate that magneto-elastic oscillations of the star
could be excited before the flare ends. Sudden readjustment of the
magnetospheric configuration could excite frequencies up to $\sim 30$
kHz, which could deliver energy to the stellar interior relatively
efficiently, though these frequencies are far higher than those of the
observed QPOs. An important question is if Alfv\'en waves have time
to excite global magneto-elastic modes before the Alfv\'en waves
damp. I stress that it is not yet known if stellar oscillations can
produce observable QPOs in the emission, though the mechanism of
\citet{timokhin_etal08} appears promising.

Realistic microphysical inputs have been used in these
analyses, but the conclusions have been drawn using models with simplified
geometries. More realistic energy deposition physics and magnetic
field geometries should be considered.

\section*{Acknowledgments}
I thank C. D'Angelo, C. Gundlach, Y. Levin, M. Lyutikov, C. Pethick,
and A. Watts for enlightening discussions. I am grateful to A. Watts
and Y. Levin for comments on the manuscript. This work was supported
by NASA Award NNX12AF88G, NWO Visitor Grant 040.11.403 (PI A. Watts),
a Monash Research Enhancement Grant (PI Y. Levin), and a Kevin
Westfold Scholarship. I thank the Astronomical Institute Anton
Pannekoek, University of Amsterdam, and Monash University, for their
hospitality.

%\bibliography{references}

\label{lastpage}

\end{document}